# Return of the Ether: Conjecture That Can Explain Photon and Electron Two-Slit Interference

Sid Deutsch
*Visiting Prof., Electrical Engineering Dept., Univ. of South Florida, Tampa*

The ether can "explain" certain questionable "quantum realities" in which single, isolated photons form a diffraction pattern in a two-slit apparatus; and, similarly, single, isolated electrons form a diffraction pattern in the equivalent of a two-slit apparatus.

The notion that every large object, such as the earth, carries its own ether, is not considered.

**INTRODUCTION**

Once upon a time, the ether belonged to physicists and electrical engineers. In 1864, James Clerk Maxwell (1831-1879) presented the equations that describe an electromagnetic field (EMF). Maxwell and his contemporaries were of the opinion that so-called empty space, or a vacuum, is really filled with a mysterious substance, "the ether." The reasoning behind the ether is this: Sound is transmitted as one molecule pushes (and, in effect, pulls by leaving a hole) its neighbor in the direction of propagation; similarly, an EMF is transmitted as one microvolume of ether pushes (and, in effect, pulls by leaving a hole) its neighbor at right angles to the direction of propagation. The velocity of sound is determined by the density and elasticity of the medium; similarly, the velocity of an EMF is determined by whatever corresponds to the density and elasticity of the ether.

But with the arrival of the twentieth century, alas, the world of James Clerk Maxwell and Isaac Newton has been modified by a considerable degree of scientific complexity. The ether has been abandoned, and quantum mechanics has changed the way we look at the atomic and subatomic world. Three of the most important tenets of quantum mechanics are:

First, energy is quantized. For an electromagnetic wave, the smallest "chunk" of energy is that of a tiny wave packet, a photon.

Second, there is a wave-particle duality. A wave, such as a photon, also behaves as if it has a certain effective mass. A particle or mass, such as an electron, also behaves as if it is a wave packet (of very high frequency, usually, because an electron has much greater mass than the effective mass of a typical photon).

Third, quantum mechanics is associated with a method for calculating *probabilities* using Schrodinger's equations. It gives the probabilities (called the *wavefunction*) of finding a photon, electron, or whatever, at a particular point in space, given the boundary probability values. The present article, however, is only concerned with the behavior of *individual* photons and electrons, for which the predictions of wavefunction probabilities may not be appropriate.

It turns out that much of quantum theory is unbelievable; that is, it runs contrary to common sense. In a photon two-slit interference pattern, a photon is presumed to be a minuscule wave packet but, if *single, isolated* photons are aimed at a double slit, the lines that build up on the recording film imply that each photon interferes with itself! It is as if the photon splits into two halves. Furthermore, although the interference pattern consists of parallel lines, the photon terminates on only a single point on the film in accordance with its probability density as a wave packet. The photon wave packet thus also behaves as a particle. The location of the single point is apparently determined by the previous history of the photon.

We can paraphrase the above with regard to an electron. A sensational experiment, another well-known example that strains quantum reality, is the electron two-slit interference pattern. An electron, presumed to be a particle, also behaves as a minuscule wave packet. If *single, isolated* electrons are aimed at a double slit, the lines that build up on the recording film imply that each electron interferes with itself! It is as if the electron splits into two halves. Furthermore, although the interference pattern consists of parallel lines, the electron terminates on only a single point on the screen in accordance with its probability density as a *wave packet*. The location of the single point is apparently determined by the previous history of the electron. Because of experimental difficulties, the "impossible" was not accomplished until 1989, when five physicists (A. Tonomura et al) used skill, persistence, ingenuity, and modern equipment to demonstrate the particle-wave duality of individual electrons.

As a result of these and other weird effects, the subject has been plagued by misgivings concerning "quantum reality." Here are some fairly recent offerings: M. Gardner, "Quantum Weirdness," in 1982; N.

Herbert, "Quantum Reality," in 1985; and D. Lindley, "Where Does the Weirdness Go?" in 1996. Several utterly nonsensical proposals have also appeared: J.S. Bell, in 1964, pointed out that, under certain experimental conditions, two photons can influence each other instantaneously, or at least at a speed much greater than that of light. H. Everett III, in 1957, proposed that there are many parallel universes; this thesis has surfaced again in D. Deutsch's "The Fabric of Reality," 1997.

The "return" of the ether is an almost obvious explanation for the weird quantum effects. We can reveal at the outset that the ether is a perfectly elastic, lossless material. That is, as a photon flies through the ether at a velocity $c$, each microvolume of ether is set into vibration, propagating the original photonic vibrations without a loss of amplitude. An electron is more complicated because it carries a charge, (which is another unknown subatomic attribute) and it moves at a velocity less than $c$, but here also there is no loss in amplitude. Both photons and electrons do not lose energy in flying through the ether; the ether does not interact with our "everyday" world where energy and mass are concerned. Photons and electrons are intimately involved with the ether, but there is never an exchange of energy or mass.

Furthermore, the ether is a "linear" material; that is, thousands of different signals (such as light and radio waves) are simultaneously superimposed everywhere around us, yet they do not interact. In a nonlinear medium, new sum and difference frequencies would be generated, but this does not happen in the putative ether. Our eyes see the light waves, undisturbed by the myriad of crisscrossing radio waves.

**THE PHOTON MODEL**

The most well-known example that strains quantum reality is a sensational experiment that involves photons. A typical EMF is the macroscopic integration of billions of photons. If the EMF is sufficiently attenuated, however, it displays its individual photons, and we can actually detect these individuals. In other words, a photon is the irreducible constituent of an EMF. It is a tiny wave packet whose frequency is the same as that of the parent EMF. A photon is a form of energy; the relation between its frequency $f$ and energy $E$ is described by $f = E/h$, where $h$ is Planck's constant. The summation of energies contained in the individual photon wave packets has to, of course, equal the energy in the "parent" EMF.

Aside from its minuscule magnitude, the main difference between a photon and a typical EMF is that the latter is a sinusoidal signal. Its energy is given in joules per second, or power in watts, whereas the energy of a photon is given in joules.

**Two-Slit Interference Pattern**

As a strategy for studying the photon, we start out, innocently enough, with the relatively strong EMF output of a laser, and then attenuate the field until individual photons can be isolated.

As a vehicle for this discussion, consider the double- or two-slit interference-diffraction apparatus of Fig. 1(a). The EMF, polarized with the <u>E</u> lines in the plane of the page, as shown, is propagating to the right. It strikes an opaque plate that contains two slits (they are at right angles to the page). Some of the EMF gets through the upper slit, and some through the lower slit. To the right of the slits, the EMF spreads out via diffraction. Two of the semi-infinite number of rays thus formed, (1) of length $\ell_1$, and (2) of length $\ell_2$, are shown as they come together on a sheet of photographic film. (Visible or ultraviolet light is usually used because their photons have sufficient energy to be recorded on the film.) What pattern will the exposed film show?

In some locations, the EMF from ray (1) is in-phase with that of ray (2) when they meet, thus increasing film exposure (constructive interference). At other locations, they have opposite phases, and the EMFs cancel (destructive interference). Figure 1(b) illustrates an in-between situation in which they are 90° out-of-phase; there is some increase in total output, by a factor of 1.414. The net results of constructive and destructive interference are the idealized set of peaks and valleys of Fig. 1(c). In Fig. 1(a), rays $\ell_1$ and $\ell_2$ are shown with relative values $b = 0.5$, $a = 10$, $y = 4$, and $\theta = 4\pi$ (because it is the second peak away from the $y = 0$ axis). The numerical values correspond to wavelength $\lambda = 0.1857$. Ray $\ell_2$ is 10.97 units long and contains 59 cycles of laser signal. Ray $\ell_1$ is 10.59 units long and contains 57 cycles. The two signals arrive in phase (constructive interference).

At $y = 1$, $\theta \cong \pi$, the longer path is 54.4 cycles long, the shorter path is 53.9 cycles long, so the difference is 0.5 cycle. The two signals cancel (destructive interference).

A change has been applied to Fig. 1(c) to convert it into the more realistic film exposure of Fig. 1(d). Since diffraction is accompanied by attenuation, Fig. 1(c) has been multiplied by a factor of the Gaussian form $\exp(-ky^2)$ to convert it into Fig. 1(d). [The accurate form of Fig. 1(d) can be derived given the various physical and wavelength dimensions, but the drawing as shown is a convenient assumption.]

Now consider that the EMF is a form of energy. If the EMF vanishes because of destructive interference, its energy must be picked up by regions of constructive interference. How this can happen is crudely shown in Fig. 2. (An exact analysis may require a high-speed computer in the hands of a thesis student, but Fig. 2 is adequate to illustrate the general idea.) We have three parallel EMFs propagating to the right: upper, middle, and lower. The numerical values give E field intensities. For some reason, such as destructive interference, the middle path is attenuated: E = 100, 80, 60, ⋯ , until it vanishes at the right. Because the E lines have to be continuous, the algebraic summation at each junction is zero. The net result is that the E lines become distorted, with substantial increases in upper and lower constructive-interference paths. (This assumes that the volume lacks charge sources.)

From the photon's point of view: A photon is a form of energy. It travels at right angles to its E (and H) lines. After each photon gets past the double slits, it diffracts by an amount that is based on its predetermined but statistically random prior experiences. Because of the bending of the E lines in Fig. 2, entering photons veer off toward the upper and lower paths, avoiding the middle destructive-interference path. When a photon strikes the photographic film, its energy is released, exposing a small dot (diameter approximately equal to the photon's wavelength).

In other words, in Fig. 1(a), the photons actually curve away from destructive-interference points $y = \pm 1$ and $\pm 3$, and toward constructive-interference points $y = 0, \pm 2$, and $\pm 4$. As a result of this "curving away," the valleys of Fig. 1(c) and (d) are created.

**Simultaneous-Burst Pattern**

Our next step is to carefully decrease the output of the laser beam. Suppose that an ideally fast shutter allows a burst of only 1000 photons to *simultaneously* fly through the slits. We are immediately faced with probabilities. Around 500 photons will probably pass through the upper slit, the remaining approximately 500 through the lower slit. Their E and H fields join up, + to -, as they disperse via diffraction. Experiments show that the film exposure display of Fig. 1(d) occurs independent of laser beam intensity. (Of course, not much, if anything, will be visible if there is a total of only 1000 photons.)

In order to refer to specific numerical values, the distribution diagram of Fig. 3(a) has been prepared. First, the $\exp(-ky^2)$ values at $y = 0, \pm 0.5, \pm 1, \cdots , \pm 5$ were added together; from the summation, a factor needed to get a total of 1000 was derived. This yielded the distribution values of Fig. 3(a). Next, each value was placed into a bin of width $y = 0.5$, as shown. The result is a crude approximation, but it is adequate for our purpose. Out of the 1000 photons, 75 will head for the $y = 0$ bin, 74 for $y = \pm 0.5$, 71 for $y = \pm 1$, and so forth. These are reasonable values, and would in fact appear as film exposure if only a single slit is open and the interference mechanism cannot operate.

The procedure used to derive Fig. 3(a) was applied to the film exposure display of Fig. 1(d), yielding Fig. 3(b). Here, out of the 1000 photons, 150 end up in the $y = 0$ bin. This is reasonable if half of the $y = 0.5$ and $y = -0.5$ photons, from Fig. 3(a), are captured by the $y = 0$ bin. But what happens to the 71 photons that, according to Fig. 3(a), start out headed for $y = 1$? Fig. 3(b) tells us that only 2 get through. What happens to the other 69 photons? They end up in the constructive-interference regions to either side of $y = 1$.

**Individual-Photon Pattern**

Finally, instead of 1000 simultaneous photons, we block the light so effectively that *only one isolated photon at a time gets through* — one per second, say. After 1000 seconds (16 2/3 minutes), we develop the film. We expect to see Fig. 3(a) because constructive or destructive interference could not possibly occur with individual photons. Instead, however,

*we get Fig. 3(b)!*

This is an unbelievable result, impossible to explain by classical physics. It defies common sense.

The evidence would have us believe that the photon somehow divides in half, and each half goes through a slit. Upon emerging from the slit, each half is apparently associated with an EMF that is similar to that of 1000 simultaneous photons (except, of course, that the total EMF energy is that of a single photon). The emerging EMFs cover the entire film of Fig. 1(a), from $y = -5$ to $+5$. The energy of the EMF that strikes the film should be modified by constructive and destructive interference, as depicted in Fig. 1(d). Instead, the photon behaves like a point particle, lands on the film at $y = 4$, say, and *all* of its energy is converted into a single bright dot at $y = 4$. After 1000 seconds, it will turn out that some 55 photons [a value given by Fig. 3(b)] were captured by the $y = 4$ bin; 150 landed in the $y = 0$ bin; and so forth.

There are two serious problems with the above recital. First, since a photon is the "irreducible constituent" of an EMF, it cannot split into two halves, each passing through one of the slits. Second, if the photon gives birth to an EMF that covers the entire film from $y = -5$ to $+5$, the photon's energy would reside in this field, leaving much less than a normal amount for the particle that eventually strikes and exposes the film at $y = 4$.

**The Wave-Particle Duality Field**

In what follows, the existence of a field that is analogous to an electromagnetic field is proposed. In the photon model of Fig. 4, it is called a Wave-Particle Duality field, or WPD field. As a more palatable example, first consider the duality field of an electron, which is a "particle" but at the same time is associated with a "wave." It usually turns out that the electron's "wave" is an X ray! However, it is an X ray in frequency only. Constructive and destructive interference patterns show that it is some kind of a field; it has a frequency that is determined by the electron's velocity when it strikes the two-slit apparatus. But it is *not* an X-ray field; exposure of the photographic film shows a single sharp point due to the electron, and not an interference pattern due to X rays (which, of course, *are* electromagnetic fields). Similarly, the WPD field of Fig. 4 is not an EMF. The drawing immediately suggests what it could be:

The conjecture is that it is a type of compression shock wave generated as the photon plows through the ether (although it is nominally a "compression" wave, it actually consists of compressions and expansions). This is analogous to air versus a high-speed projectile. Air supports the propagation of sound waves, and a projectile forms a shock wave. The shock wave consists of compressions (and expansions) propagating at the speed of sound. Constructive and destructive interference always shows up when the shock wave reaches a reflecting object or refractive medium.

Analogously, the ether supports the propagation of EMFs, and the photon "projectile" forms a shock wave that propagates at the same speed. It would be premature, however, to think that the WPD field really is a shock wave. We know a great deal about air and sound shock waves, but we do not know what the ether or an electric or magnetic field really are. Despite this ignorance, we get through life drawing electric and magnetic field lines, and designing sophisticated equipment based upon imaginary field intensities and flux densities. In the world of imaginary field lines that follows, we assume that the WPD field lines of Fig. 4 really exist because they are associated with constructive and destructive interference.

However, one should not pursue the analogies too far. A sonic boom carries a tremendous amount of energy, but the WPD field may not carry any energy at all. Zero energy? The ether is a peculiar medium: we peer at photons, tiny wave packets that have been traveling for billions of years through the ether *without attenuation*. From another viewpoint, there can be no attenuation because the latter implies the conversion of photon energy into heat, which in turn implies that some particle that has mass (such as an atom) will vibrate more rapidly as it absorbs this energy. But there are no atoms in the ether, or at least none that has absorbed the energy of this billion-year-old wave packet (which is why we can detect it, of course). In other words, the ether is a perfectly elastic medium; the transverse ripple is passed along, without change, at the velocity of propagation.

Closer to home, and something about which we know a great deal, there is the zero attenuation of superconductivity and superfluidity:

For many electrical conductors (and, recently, semiconductors), if they are cooled towards 0 K, a transition temperature is reached at which, suddenly, electrical resistance vanishes. Other changes also take place at the transition temperature: magnetic fields are expelled, and thermal properties are altered.

Helium liquifies at 4.22 K. If it is cooled to 2.172 K, a transition occurs at which, suddenly, viscosity vanishes. The superfluid is able to flow at high speed through tiny holes. Here, also, other changes take place at the transition temperature.

Before the days of superconductivity and superfluidity, we could not conceive of zero electrical resistance and zero viscosity. They were amazing experimental discoveries. In this same spirit of open-mindedness, we may conjecture that the WPD field can certainly be a zero-energy field if it is not required to do work. From here on, in this article, it is conjectured that the WPD field shock wave consists of compressions (and expansions) of the ether that, somehow, do not convey any energy.

Figure 4 is of course meant to be a schematic representation. Inside the "power pack" is a wave packet, a minuscule EMF whose frequency is the most important specification of the photon, with energy given by $E = fh$. In the side view, the photon is flying off to the right at velocity $c$. Preceding it is the WPD field. Lines in the so-called compression shock wave have the same spacing as those of the power pack because the frequencies are the same; this is shown to be so by interference patterns.

The WPD field is polarized as depicted in the end view. (Vertical polarization is shown, so this is also the direction of the E lines in the power pack.)

The WPD field extends over a cone whose projection, in Fig. 4, runs from +45° to -45° relative to the axis of propagation. The ±45° angle is a matter of convenience, and is much more than is necessary to demonstrate interference in an actual apparatus.

How far does the WPD field extend in front of the power pack? At least 10 or 20 wavelengths, enough to get a reasonably effective degree of destructive interference. The WPD field may therefore be finite, like the strong-force field. It may be infinite, like an $\underline{E}$ or $\underline{H}$ or graviton field, but this is unlikely because it would violate the "maximum velocity = $c$" rule: If the power pack is moving to the right with velocity $c$, and the WPD field expands toward infinity with velocity $c$, then the WPD field would move with velocity $2c$ relative to a stationary reference. Following this line of thought, it is more reasonable to conjecture that a hydrogen electron, as it spirals from the $n = 2$ to $n = 1$ orbit, first generates a finite WPD field before it releases the photon's power pack. The electron spiral could be a very gradual multi-revolution locus.

Views (a) to (g) of Fig. 5 depict how the photon WPD model of Fig. 4 can explain the single, isolated-photon two-slit experimental results of Fig. 1 (the slits are greatly magnified for the sake of clarity):

In (a), the photon is approaching the two-slit apparatus.

In (b), the leading portion of the WPD field has split, with a fragment getting through each of the slits. The fragments diffract. Thus far, the action is identical to that of a laser beam directed at the two slits.

In (c), we depart from a conventional perspective. The power pack and at least some of the WPD field are inseparable, since it is impossible to generate a shock wave without the power pack. In getting to (c), the photon has three choices: 1)Strike the slit plate at the center, in which event the power-pack's energy is converted into heat, and the WPD field vanishes without a trace; 2)The power pack can pass through the upper slit. This is the choice shown in Fig. 5(c); 3)The power pack can pass through the lower slit. The actual path taken by the photon is predetermined but statistically random, based upon its prior history.

There is a serious problem here with regard to the lateral movement from (b) to (c). Since the photon has zero mass, one could think that it can be pushed sideways without the expenditure of force. This is not so for a photon, which travels at the speed of light. The *effective* mass is given by $E = mc^2$ or, combined with $E = fh$, we have $m_{eff} = fh/c^2$.

As a numerical example: For the photon generated when an electron spirals from the $n = 2$ to $n = 1$ orbit of a hydrogen atom, $f = 2.467 \times 10^{15}$ Hz, so we get $m_{eff} = 1.819 \times 10^{-35}$ kg. This is a truly minuscule mass. It is 50,000 times lighter than an electron. An effective photon mass that equals that of an electron is obtained with a frequency of $1.236 \times 10^{20}$ Hz; this is on the borderline between X and gamma rays.

Nevertheless, despite its minuscule effective mass, a finite force has to act on the photon to achieve lateral deflection. If the two-slit experiment is performed using a laser beam, there is plenty of energy in the EMF to support lateral movement, but not with a single, isolated photon.

Although it may not be valid to think of the photon as being similar to a high-speed projectile in air, the analogy suggests a solution to the lateral-force problem. The conjecture is that the ether forms stream lines through the two slits, and these guide or steer the photon. The ether supplies the lateral force, much as a glancing blow can force a projectile in air to change its course. If no change in speed is involved, the lateral push need not entail a change in energy.

The lateral force is reminiscent of the force of attraction between two conducting, uncharged plates brought sufficiently close together in a high vacuum. The minuscule force is known as the H.B.G. Casimir effect. The force, which has been measured [S.K. Lamoreaux, 1997], is another zero-energy phenomenon.

What are stream lines? In smoothly flowing water (a non-turbulent "stream"), they trace out the flow lines. Think of the ether as flowing through the slits. This implies that the ether is not a passive jelly. The conjecture here is that the ether is a perfectly elastic medium in which stream lines are ubiquitous. The stream lines in an all-pervading ether guide the compression shock waves; this is reminiscent of the pilot wave proposal of David Bohm (1919-1992).

Returning to Fig. 5: (d) is the same as (c), except that the WPD fields are omitted for the sake of clarity. We now see that the particular WPD field fragment to which the power pack was attached, in (c), has directed the power pack to $y = 3$.

In (e), the power pack is midway between the two-slit plate and the photographic film. Because it is approaching a destructive-interference point, the WPD field lines are concave, as in Fig. 2. This translates into ether stream lines that laterally push or "encourage" the power pack to head for the constructive-interference points at $y = 2$ or 4.

In (f), the power pack is shown on a path toward $y = 4$.

In (g), the power pack arrives at the film, exposing a tiny dot at the $y = 4$ position. According to Fig. 3(b), if 1000 individual photons are launched in this way, in sequence, 55 of them will end up in the $y = 4$ slot, and only 2 in the $y = 3$ slot.

Figure 5(g) shows the path taken by the power pack. The various curves are explained as the result of lateral forces exerted by the ether upon the photon. The WPD field is an ethereal compression shock wave; it vanishes without a trace.

The concept that the WPD field may be a longitudinal wave is depicted in Fig. 6(a). To the right of the power packs, black and white strips symbolize compression and expansion of the ether, respectively. The split paths, in which the power pack proceeds through the upper slit, is illustrated.

In Fig. 6(b), the WPD field lines interfere; the ethereal stream lines follow the interference maximum summation peaks. These correspond to regions where the E field intensity is maximum in Fig. 2; it is these points that guide the stream lines, which "encourage" the power packs to end up near constructive interference maximum points.

**Decaying-Exponential WPD Field**

A more detailed consideration of the photon's wave-particle duality field, assuming that it is a decaying exponential, is depicted in Fig. 7(a). This is strictly a viewpoint taken from classical physics. Because the shape of the field and its spectrum should be independent of frequency, a convenient spatial frequency of 1 cycle/meter is used in the figure. An exponential length constant of 10 cycles is shown; that is, after 10 cycles, the amplitude is $\varepsilon^{-1} = 0.368$ relative to its value at $x = 0$. This is accompanied by a reasonably high-$Q$ spectrum in (b).

Figure 7(a) shows a *spatial* waveform rather than a time waveform. The time waveform is seen by a stationary observer as the photon flies by. It is Fig. 7(a) reversed; that is, a rising exponential. It is a matter of convenience as to which waveform is used. The idea in Fig. 7 is that the WPD field has to be a good sine wave in order to get good destructive interference in the two-slit experiment.

The magnitude of the spectrum of Fig. 7(a) is plotted in Fig. 7(b), with $f = \omega/2\pi$. This is a *spatial frequency* spectrum in cycles/m rather than cycles/s. All-important is the quality factor, or $Q$. The –3 dB (0.707 peak) level in Fig. 7(b) extends from $f = 0.9838$ to $1.0157$, so $Q = 1/0.0319 = 31$. This is a reasonably high value.

There are indications that the $Q$ does not remain constant. According to Raymond Y. Chiao et al, the photon wave packet (the power pack in Fig. 4) becomes shorter if it passes through a transparent barrier, such as glass, that slows it down. Upon emerging from the barrier, the speed returns to normal ($c$ in vacuum or air), but the shorter wave packet can be interpreted as a steeper exponential decay. This may correspond to increased bandwidth and lower $Q$ in the putative WPD field.

**THE ELECTRON MODEL**

Another sensational experiment that strains quantum reality is one that involves electrons. One of the problems here is that objects that have mass, such as electrons, become heavier (and shorter) as their velocity increases. Increases, that is, relative to the stationary, non-accelerating observer who is making the measurements. Therefore, the changes in effective mass and length due to relative velocity are called *relativistic*.

The relativistic change in length (the Lorentz contraction) is not pertinent to the present discussion; only the relative change in mass is considered. For convenience, although we are concerned here with objects that have mass, only the electron rather than proton or neutron is considered. Much of the discussion and conclusions, however, also apply to the proton and neutron.

"Massive" particles display gravitational attraction toward each other. If massive particles interact, momentum = mass × velocity is conserved. It is interesting to contrast this with the "massless" photon: Photons ignore each other, and two photons that hit each other head-on only yield the algebraic sum of their respective wave packets. Following the "collision," they continue to propagate, unchanged, at the speed of light.

A permanent record of electron strikes can be obtained by placing a photographic film next to the fluorescent screen of a cathode-ray tube. Here, the potential energy of the electric field is converted into kinetic energy. We should use the accurate relativistic form for velocity $v$, especially since it is tractable. An electron *behaves* as if it has an effective mass given by $m_{eff} = \gamma m_0$, where $\gamma$ is the relativistic increase in mass factor. Various values, as a function of $V$, are given in Table 1.

A photon, which is an electromagnetic wave packet, displays the characteristics of a particle that has mass. In 1924, Louis de Broglie (1892-1987) proposed that the reverse may be true — that an electron, which has mass, can display the characteristics of a wave. Soon afterwards, experiments showed that de Broglie's hypothesis was correct; in fact, every mass in motion, in general, demonstrates wave characteristics.

The reason for considering the massive electron versus the wavelike photon is that each of them displays an interference pattern in the two-slit apparatus. They are, however, two different species: The electron's field travels at the speed of the electron, which can be anything from zero up to the upper limit, the speed of light, while the photon's WPD field *always* travels at the speed of light. Also, the frequency of the electron's field is a function of its velocity, while the photon's WPD field frequency is that of its power pack. Therefore, in what follows, the electron's field is called a particle-wave duality (PWD) field to distinguish it from the photon's WPD field.

The PWD frequency of an electron is given by the photon's $f = m_{eff}c^2/h$ with the substitution of the electron's velocity $v$ in place of the photon's velocity $c$. This yields $f = \gamma m_0 v^2/h$. This is the equation used to calculate values in the frequency column of Table 1. Wavelength is given by $v/f$, or $\lambda = h/(\gamma m_0 v)$.

**Two-Slit Interference Pattern**

The frequency values in Table 1 are relatively high. As mentioned above, an electron is a giant compared to a photon, and this shows up in the associated frequency values. At a typical cathode-ray tube value of $V = 25{,}000$ volts, Table 1 shows $f = 1.181 \times 10^{19}$ Hz. This is an X-ray frequency, as is also indicated in the last column of Table 1. We hasten to add that these are *not* the X rays that, it is frequently claimed, are emitted by a cathode-ray tube. The electron's particle-wave dual is an X ray in frequency only; it is not an electromagnetic wave; it propagates at the velocity of the electron, not that of light; it has zero energy, zero penetrating power, and vanishes without a trace when the electron strikes its fluorescent screen. Is it realistic for us to believe that it has zero energy? The arguments regarding energy of the photon's zero-energy WPD field apply equally well to the electron's PWD field.

The bona fide X rays that the screen emits are due to the great velocity with which an electron arrives at the screen. Part of the kinetic energy is converted into fluorescent excitation, part into photons in the X-ray range of frequencies, and part into heat. In the case of a television receiver, it is generally considered that the X-ray effect is negligibly small, especially compared to that of deadly program material.

Nevertheless, the high PWD frequencies offer almost insurmountable experimental difficulties in the attempt to demonstrate the incontestable signature of a wave — constructive and destructive interference in the two-slit apparatus. It is interesting to consider, below, how some of the difficulties were overcome.

The proof that an electron can act as a wave came from the same techniques that are used to prove that an X ray is a wave. For example, the above-mentioned $V = 25{,}000$-volt PWD frequency has a wavelength of 0.077 angstrom. In Fig. 1, the spacing between the two slits is around 5 wavelengths, so a spacing of 0.4 angstrom would be reasonable for the electron beam. The "slits" in this case can be provided — many of them — by the repetitive spacing between the atoms of a crystalline material. Clinton Davisson and Lester Germer, in 1925, showed electron diffraction and interference using a crystal made out of nickel.

In 1989, as previously cited, A. Tonomura et al demonstrated the particle-wave duality of electrons. In what follows, advantage will be taken of the accomplishment of Tonomura et al by using the two-slit interference drawings of the photon and applying them to two-slit interference of the electron. Changes in text and drawings are made, as needed, to accommodate electrons rather than photons.

As an electron source, Tonomura et al used a sharp field-emission tip and an anode potential of 50 kV. According to Table 1, $f$ and $\lambda$ were $2.3 \times 10^{19}$ Hz and 0.054 angstrom. From Tonomura et al, "When a 50-kV electron hits the fluorescent film, approximately 500 photons are produced from the spot." They used a much more sophisticated light-gathering arrangement, including a magnification of 2000, than a photographic film placed next to a fluorescent screen.

For electrons, one must employ a high vacuum, in addition to facing the problems associated with angstrom-size wavelengths. As a vehicle for this discussion, consider the idealized two-slit interference-diffraction apparatus of Fig. 1, *with the laser beam replaced by an electron beam, and a fluorescent screen added in front of the photographic film*.

The electron beam is moving to the right. It strikes a plate that contains two slits. Some of the electrons get through the upper slit, and some through the lower slit. To the right of the slits, the electrons spread out, via diffraction, as if they had wave characteristics. Two of the semi-infinite number or rays thus formed, (1) of length $\ell_1$ and (2) of length $\ell_2$, are shown as they come together on a fluorescent screen. A relatively high voltage is used so that the electrons will have sufficient energy to elicit a fluorescent response that can be recorded on the film. What pattern will the exposed film show? It will show a pattern corresponding to Fig. 1(d).

Now consider that the electron beam carries kinetic energy. If an electron does not arrive at the screen because of destructive interference, it must be picked up by regions of constructive interference. How this can happen is crudely shown in Fig. 2. We now have three parallel PWD fields propagating to the right: upper, middle, and lower. The numerical values give putative electric field intensities if the PWD fields are analogous to electromagnetic fields; although they are *not* EMFs, but behave like EMFs, we speculate that we can use the analogy.

After each electron gets past the double slits, it diffracts by an amount that is based on its predetermined but statistically random prior experiences. Then, because of the bending of the pseudo-E lines in Fig. 2, entering electrons veer off toward the upper and lower paths, avoiding the middle destructive-interference path. In other words, in Fig. 1(a), the electrons actually curve away from destructive-interference points $y = \pm 1$ and $\pm 3$, and toward constructive-interference points $y = 0, \pm 2$, and $\pm 4$. As a result of this "curving away," the valleys of Fig. 1(c) and (d) are created.

Our next step is to carefully decrease the output of the electron beam. Suppose that an ideally fast pulse allows a burst of only 1000 electrons to *simultaneously* fly through the slits. The procedure used to derive Fig. 3(b) for a photon beam again results in Fig. 3(b). Here, out of the 1000 electrons, 150 end up in the $y = 0$ bin, and so forth.

**Individual-Electron Pattern**

Finally, instead of 1000 simultaneous electrons, we restrict the beam so effectively that *only one isolated electron at a time gets through* — one per second, say. After 1000 seconds, we develop the film. We expect to see Fig. 3(a) because constructive or destructive interference could not possibly occur with individual electrons. Instead, we get Fig. 3(b). This is an unbelievable result, impossible to explain by classical physics.

The Tonomura paper is titled "Demonstration of Single-Electron Buildup of an Interference Pattern." In its entirety their Abstract follows: "The wave-particle duality of electrons was demonstrated in a kind of two-slit interference experiment using an electron microscope equipped with an electron biprism and a position-sensitive electron-counting system. Such an experiment has been regarded as a pure thought experiment that can never be realized. This article reports an experiment that successfully recorded the actual buildup process of the interference pattern with a series of incoming single electrons in the form of a movie."

The Tonomura et al experiments show that the film exposure display of Fig. 1(d) occurs independent of electron beam density. Their paper reproduces five film exposures showing how the electron interference pattern builds up as the number of *individual* electrons striking the fluorescent screen increases.

The evidence would have us believe that an electron somehow divides in half, and each half goes through a slit. Upon emerging from the slit, each half is apparently associated with an EMF that is similar to that of 1000 simultaneous electrons (except that the total EMF energy is that of a single electron). The emerging EMFs cover the entire screen of Fig. 1(a), from $y = -5$ to $+5$. The energy of the EMF that strikes the screen should be modified by constructive and destructive interference, as depicted in Fig. 1(d). Instead, the electron behaves like a point particle, lands on the screen at $y = 4$, say, and *all* of its energy is converted into a single bright dot at $y = 4$. After 1000 seconds, it will turn out that some 55 electrons [a value given by Fig. 3(b)] were captured by the $y = 4$ bin; 150 landed in the $y = 0$ bin; and so forth.

There are two serious problems with the above recital. First, since an electron is an irreducible constituent of matter, it cannot split into two halves, each passing through one of the slits. Second, if the electron gives birth to an EMF-type field that covers the entire screen from $y = -5$ to $+5$, the electron's energy would reside in this field, leaving less than a normal amount for the particle that eventually strikes and stimulates fluorescence at $y = 4$.

**The Particle-Wave Duality Field**

In what follows, it is proposed that the electron is accompanied by a PWD field, depicted in Fig. 8, that is similar to the photon's WPD field of Fig. 4. "Similar," but different in two major respects: 1)The electron and its entourage can move at any velocity *less than c*, whereas a photon propagates at velocity *c* (through a vacuum); 2)The frequency of the electron's PWD field is a function of *v*, whereas the frequency of the photon's field is equal to that of its power pack (the wave packet).

The conjecture here is that the electron's PWD field is a type of compression *wind* generated as the electron flies through the ether (although it is nominally a "compression" wind, it actually consists of compressions and expansions). This is analogous to air versus a low-speed projectile, such as a pitched baseball. In the ether, the PWD field of Fig. 8 corresponds to compressions and expansions that precede the power pack. It is again conjectured that these ethereal waves do not convey any energy.

Figure 8 is meant to be a schematic representation. Inside the "power pack" is a negative charge, $e = 1.602 \times 10^{-19}$ C, mass $m_0 = 9.109 \times 10^{-31}$ kg, and normalized spin $s = ½$. The spin of a particle is its angular momentum that exists even when the particle is at rest, just as it has a mass $m_0$ at rest. In the side view, the electron is flying off to the right at velocity *v*. Preceding it is the PWD field whose frequency and wavelength are given, for various values of *V*, in Table 1.

The picture that emerges is this: An electron at rest has a negative charge *e*, mass $m_0$, and spin *s*. As soon as it starts to move, a PWD field develops. For example, when it has converted 1 volt into kinetic energy, Table 1 tells us that the electron model is moving (to the right, say) at a velocity of 593,000 m/s. This is relatively slow for an electron. The PWD field lines are 6 angstroms apart between + and – (the wavelength is 12 angstroms). The lines zoom by at a frequency of $4.836 \times 10^{14}$ Hz. Although this corresponds to an orange glow, there is of course no visible effect when the electron strikes the two-slit plate. The PWD field, to repeat, is not an EMF, and probably carries zero energy.

As an electron accelerates, frequency increases and the wavelength shrinks. Relativistic effects become appreciable; the electron behaves as if its mass is increasing in accordance with $\gamma m_0$. At a potential of 510,990 volts, $\gamma = 2$, the PWD frequency is $1.85 \times 10^{20}$ Hz, and $\lambda = 0.014$ angstrom.

The experiment of Tonomura et al shows that the PWD really exists. It may not look like the wave peaks depicted in Fig. 8, but the electron interference pattern is there, literally in black and white. The pattern agrees with the 50,000-volt calculated wavelength of 0.054 angstrom. At this voltage, relativistic effects are also verified since $\gamma$ is appreciably greater than 1 (it is 1.10).

Is the PWD field longitudinal, like a sound wave, or transverse? For a photon, polarization shows that the WPD field is transverse. In "copycat" fashion, therefore, the electron's PWD field is shown as having a polarization plane in Fig. 8, but this is conjecture. If the field is a compression wind wave in the ether, however, it is analogous to a longitudinal wind disturbance in air, and the polarization plane becomes meaningless.

The changes in effective mass and PWD wavelength occur because the electron is moving — with respect to what? With respect to the electron gun in a cathode-ray tube in a physics laboratory? Why should movement induce the mass change? What about the relativistic effect: are we prepared to say that an observer moving with the electron (as it drifts at constant speed past the anode, say), will see no change in mass and no PWD field? It may be much easier to visualize these changes if an ether is present. As the electron flies through the ether, a "viscosity" interaction induces wind waves (the PWD field) and also resists high particle speed via an effective increase in mass. The latter could simply be due to the effective mass of the ether that is carried along by the power pack.

So a photon can travel through the ether at the speed of light, without attenuation, whereas an electron runs up against an ether that has effective mass. These are, indeed, strange conjectures.

Views (a) to (g) of Fig. 5 now depict how the electron model of Fig. 8 can explain the single, isolated-electron two-slit experimental results of Fig. 1. The text would follow almost word-for-word the photon discussion in connection with Fig. 5.

With regard to the lateral force needed to change the electron's direction in going from Fig. 5(b) to (c), it is again conjectured that the ether forms stream lines through the two slits. These guide or steer the electron as the ether supplies the lateral force that is required. If no change in speed is involved, the lateral push need not entail a change in energy.

In Fig. 5(e), the power pack is midway between the two-slit plate and the fluorescent screen. Because it is approaching a destructive-interference point, the PWD field lines are concave, as in Fig. 2. This "encourages" the power pack to head for the constructive-interference points at $y = 2$ or 4.

Figure 5(g) shows the path taken by the power pack. The PWD field is an ethereal compression wind; it vanishes without a trace.

Tonomura et al do not attempt to explain the unrealistic experimental outcome. The statistical predictions of quantum mechanics are of no help here because we are dealing with the interference pattern associated with a *single* electron.

**Decaying-Exponential PWD Field**

A more detailed consideration of the electron's PWD field, assuming that it is a decaying exponential, is depicted in Fig. 7(a), except that the "power pack" is that of the electron of Fig. 8. The text would follow almost word-for-word the photon discussion in connection with Fig. 7. The idea in Fig. 7 is that the PWD field has to be a good sine wave in order to get good destructive interference in the two-slit experiment.

The spectrum of Fig. 7(a) is plotted in Fig. 7(b), with $f = \omega/2\pi$. This is a *spatial frequency* spectrum in cycles/m rather than cycles/s. All-important is the quality factor, or $Q$, which has a reasonably high value, 31.

**CONCLUSIONS**

The ether, if it exists, can "explain" certain questionable "quantum realities." The ether is a perfectly elastic, lossless, linear "material."

(1) Single, isolated photons form a diffraction pattern in a two-slit apparatus. It is as if the photon splits into two halves. The photon is associated with a wave-particle duality (WPD) field. The ether conjecture is that a zero-energy shock wave is formed, consisting of etherial compressions (and expansions). It is the shock wave that interferes with itself in the apparatus. The ether laterally guides the photon, which has an effective mass, via stream lines.

(2) Repeat (1) for a single, isolated electron. The electron is associated with a particle-wave duality (PWD) field. Because electron velocities may be relatively slow, the "shock" wave is called a "wind" wave.

The case for resuscitation of the ether is based on the behavior of individual photons and electrons. The notion that every large object, such as the earth, carries its own ether, has not been considered.

**Table 1.** Various values associated with an electron as potential energy $eV$ is converted into kinetic energy $K$. $\gamma$ is the relativistic increase in mass factor; $f_{PWD}$ and $\lambda_{PWD}$ are frequency and wavelength of the particle-wave duality field. Because this is *not* an electromagnetic field, the last column is for identification only; no Orange, UltraViolet, X ray, or Gamma ray energy is actually available.

| $V$, volts | $\gamma$ | $v/c$ | $v \times 10^8$, m/s | $f_{PWD}$, Hz | $\lambda_{PWD}$, angstroms | ID |
|---|---|---|---|---|---|---|
| 1 | 1.000 | 0.00198 | 0.00593 | $4.836 \times 10^{14}$ | 12.26 | O |
| 10 | 1.000 | 0.00626 | 0.01876 | $4.836 \times 10^{15}$ | 3.878 | UV |
| 100 | 1.000 | 0.01978 | 0.05930 | $4.836 \times 10^{16}$ | 1.226 | UV |
| 1000 | 1.002 | 0.06247 | 0.1873 | $4.831 \times 10^{17}$ | 0.3876 | X |
| 10000 | 1.020 | 0.1950 | 0.5846 | $4.790 \times 10^{18}$ | 0.1220 | X |
| 25000 | 1.049 | 0.3018 | 0.9049 | $1.181 \times 10^{19}$ | 0.07664 | X |
| 50000 | 1.098 | 0.4127 | 1.237 | $2.310 \times 10^{19}$ | 0.05355 | X |
| 100000 | 1.196 | 0.5482 | 1.644 | $4.440 \times 10^{19}$ | 0.03701 | X |
| 510990 | 2 | 0.8660 | 2.596 | $1.853 \times 10^{20}$ | 0.01401 | $\gamma$ |
| $10^6$ | 2.957 | 0.9411 | 2.821 | $3.236 \times 10^{20}$ | 0.008719 | $\gamma$ |
| $10^7$ | 20.57 | 0.9988 | 2.994 | $2.536 \times 10^{21}$ | 0.001181 | $\gamma$ |
| $10^8$ | 196.7 | 1.0000 | 2.998 | $2.430 \times 10^{22}$ | 0.000123 | $\gamma$ |

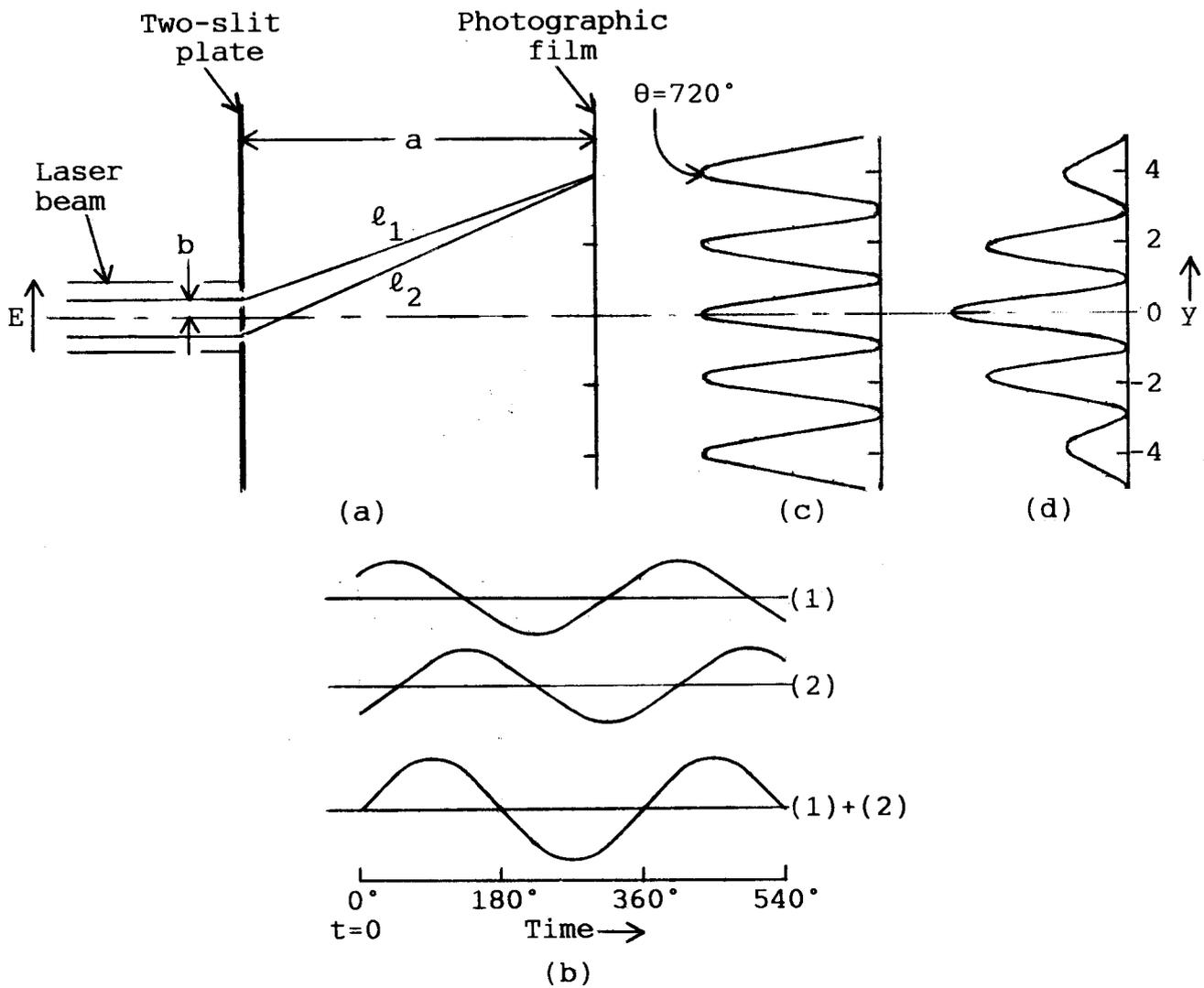

**Fig. 1.** Two-slit interference and diffraction. (a)Schematic of apparatus. The slits are at right angles to the page. Two of the rays leaving the slits are depicted as they meet at $y =$ 4 of the photographic film. (b)Waveforms of rays (1) and (2) when they meet at the film if they are 90° out of phase. (c)Idealized film pattern. (d)The film pattern "corrected" by adding the attenuation that accompanies diffraction.

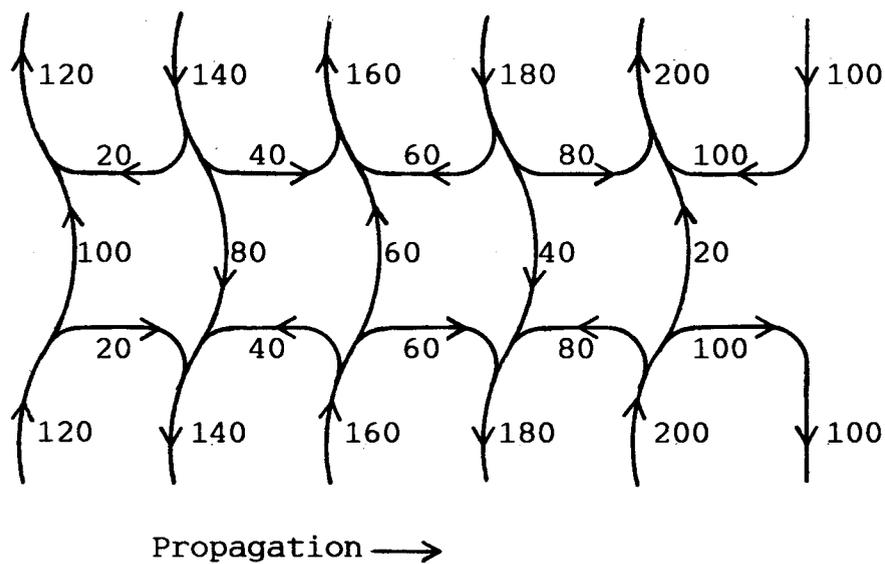

**Propagation ⟶**

**Fig. 2.** Crude depiction of <u>E</u> field intensities if EMF is propagating to the right, with destructive interference in the middle path, but with constructive interference in the upper and lower paths. Because of the interaction between concave and convex lines, photons are directed toward regions of constructive interference.

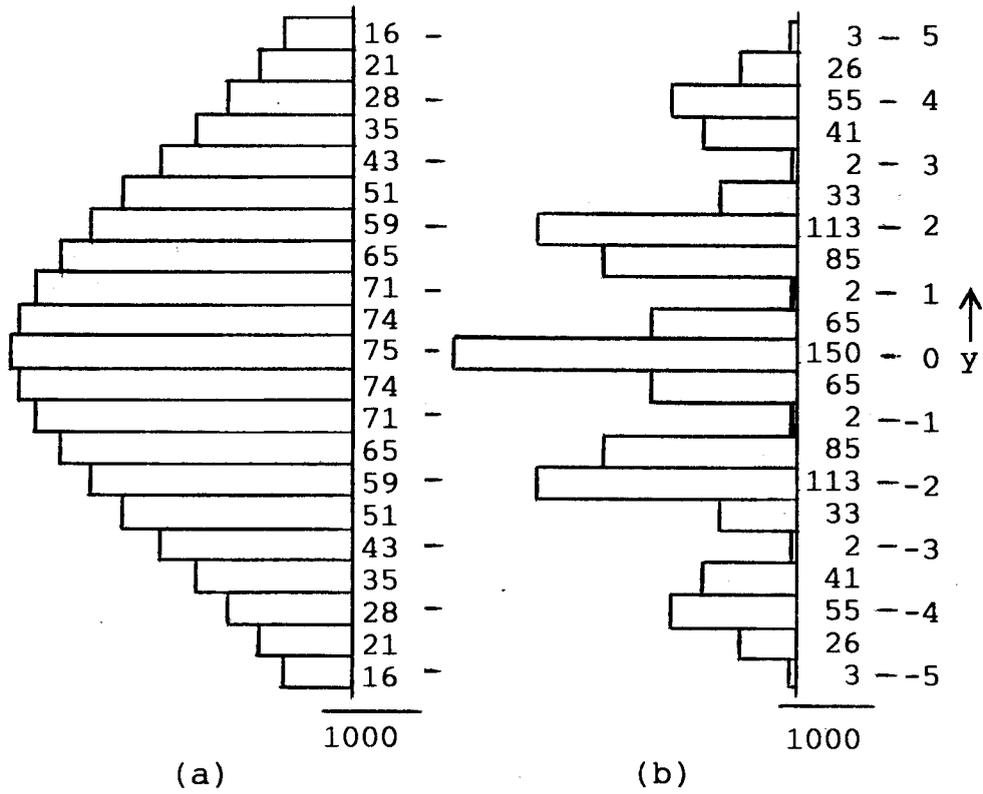

Fig. 3. Photon exposure distributions at the film of Fig. 1(a) if bins are $0.5y$ unit wide. (a)Due to an assumed diffraction attenuation function, $\exp(-0.0625y^2)$, with interference effects omitted. (b)Including constructive and destructive interference, as in Fig. 1(d).

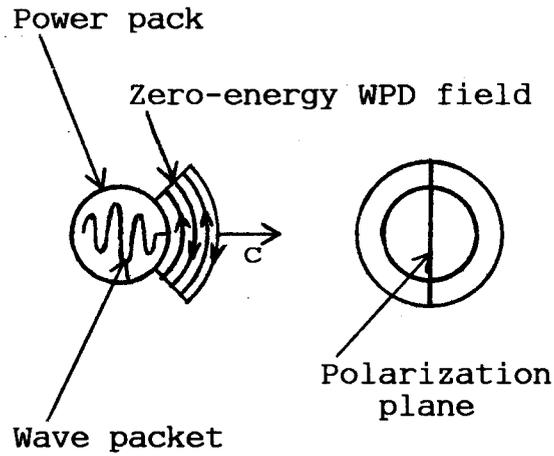

**Fig. 4.** Schematic model of a photon that can account for single, isolated-photon two-slit interference effects. The power pack contains EMF wave packet energy, $E = fh$. It is preceded by a zero-energy wave-particle duality (WPD) field as the photon moves to the right with velocity $c$. The WPD field may simply be a type of compression shock wave generated as the photon plows through the ether.

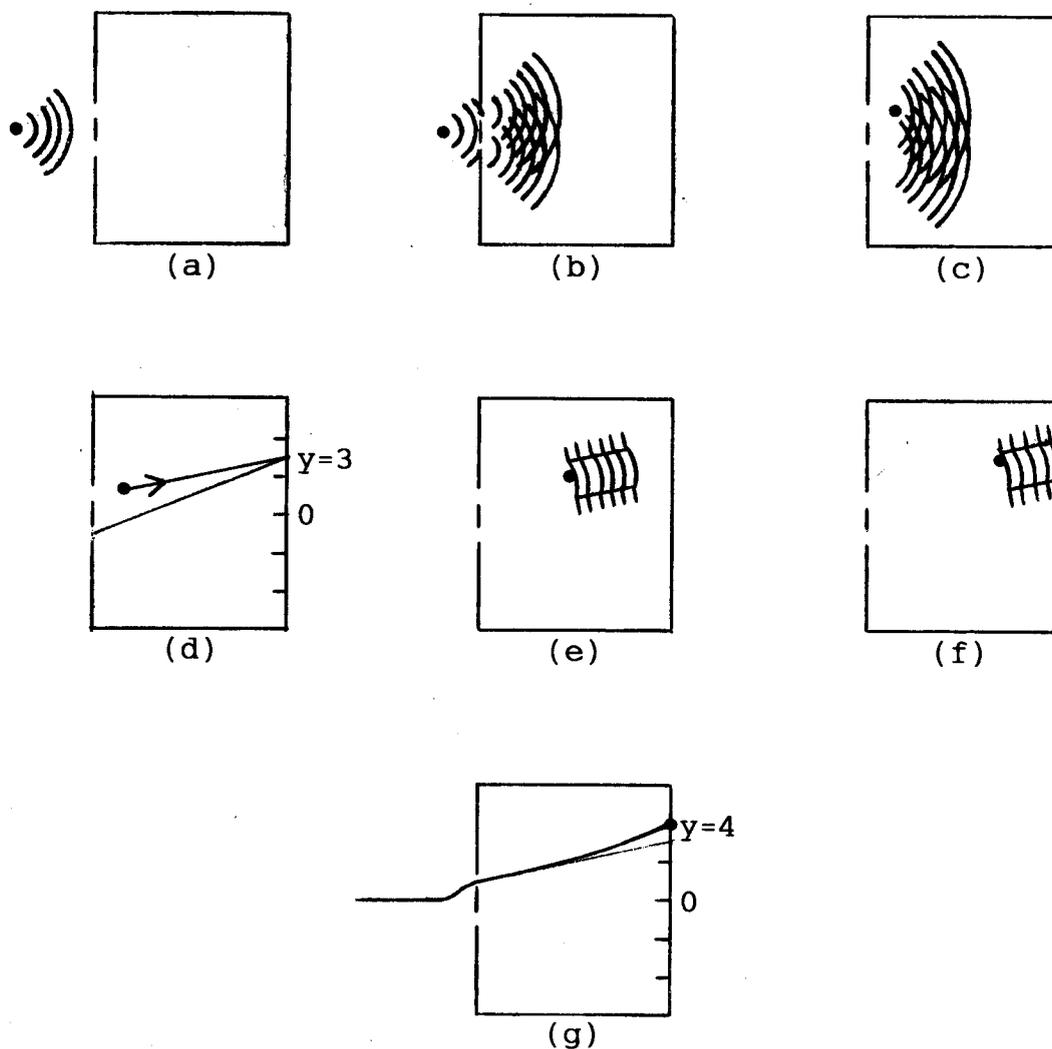

Fig. 5. Sequence that illustrates two-slit interference effects that accompany a single, isolated photon. (a)Photon approaching the slit plate. (b)Leading portion of WPD field has split, with a fragment getting through each of the slits. (c)The WPD fields have progressed beyond the slit plate. The power pack, because of predetermined but statistically random past history, has followed the upper-slit WPD segment. (d)Same as (c), but with WPD fields omitted. The power pack is heading for the $y = 3$ point of the photographic film. (e)The power pack and net WPD field, halfway across. (f)Because WPD field lines are concave, the power pack is directed away from the destructive-interference $y = 3$ point. (g)The power pack locus curves, exposing film at the $y = 4$ point. The ethereal WPD field has vanished without a trace.

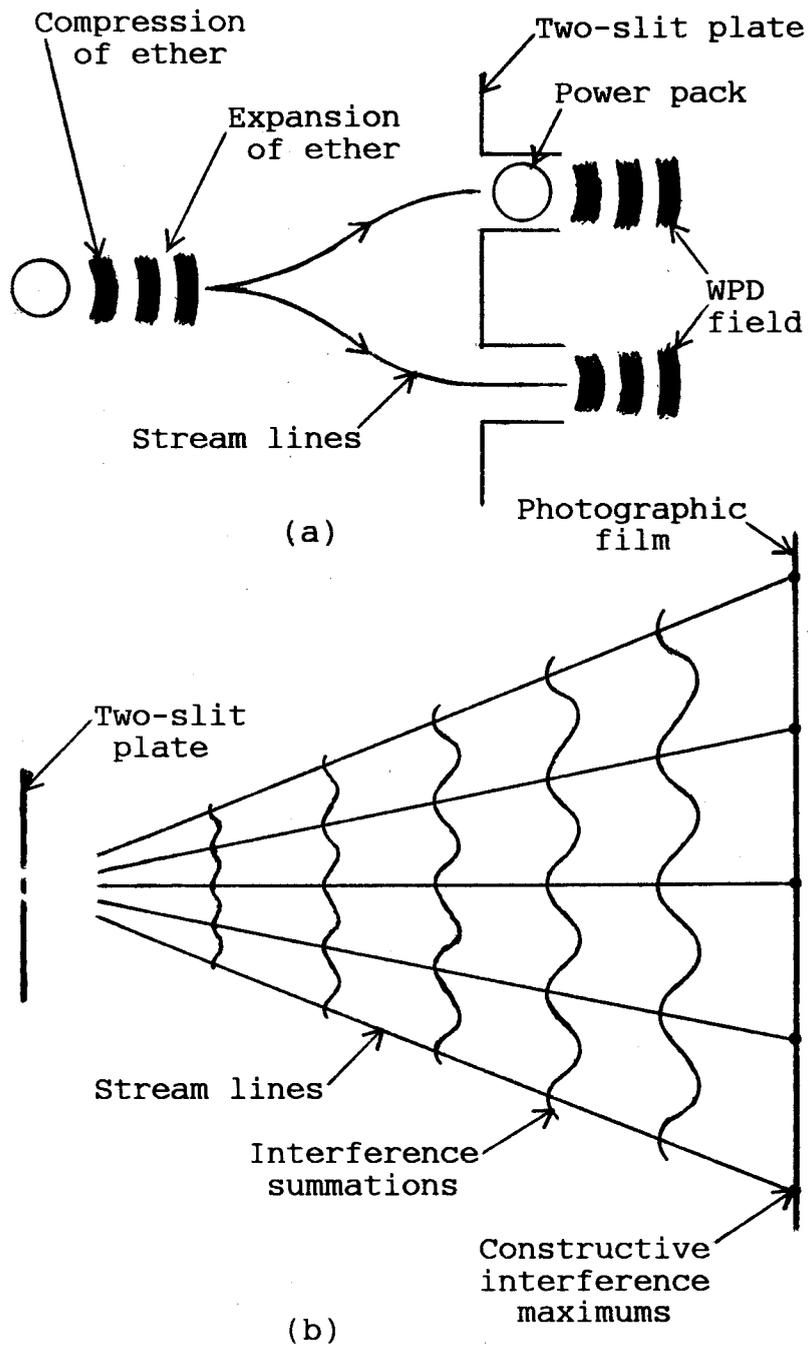

**Fig. 6.** Additional illustration of two-slit plate interference. (a)The WPD field is depicted as a longitudinal wave. (b)The ethereal stream lines follow the interference maximum summation peaks.

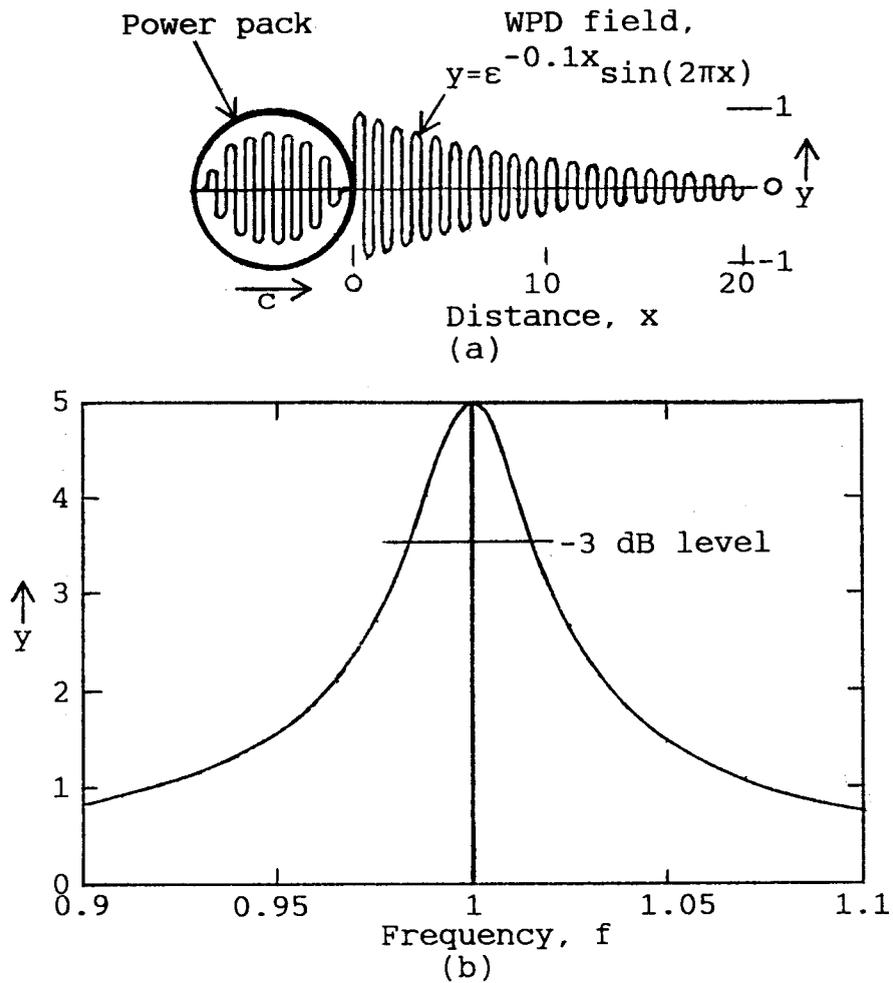

**Fig. 7.** A more detailed view of the photon's WPD field, assuming that it is a decaying exponential. (a)Power pack and WPD field moving to the right with velocity $c$. The spatial frequency is 1 cycle/meter. (b)Spatial frequency spectrum of the WPD field, in cycles/meter. The $Q$ is 31.

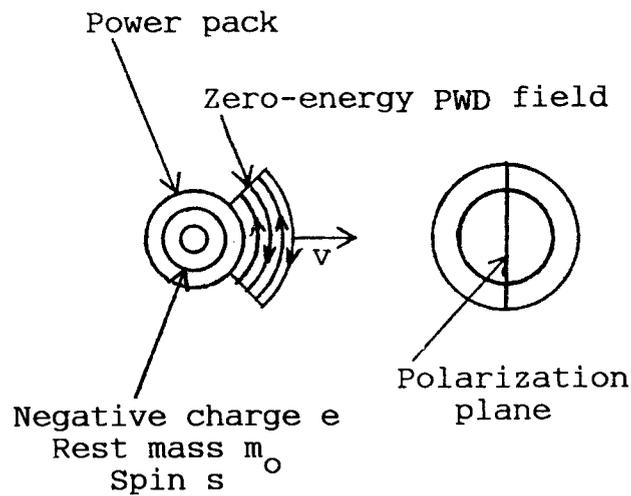

**Fig. 8.** Schematic model of an electron that can account for single, isolated-electron two-slit interference effects. The power pack represents charge, rest mass, and spin. It is preceded by a zero-energy PWD field as the electron moves to the right with velocity $v$. The PWD field may simply be a type of compression wind generated as the electron flies through the ether.